\documentclass[conference]{IEEEtran}
\IEEEoverridecommandlockouts
\usepackage{cite}
\usepackage{amsmath,amssymb,amsfonts}
\usepackage{graphicx}
\usepackage{xcolor}
\usepackage{dblfloatfix}
\usepackage{tikz}
\usepackage{url}
\usepackage{balance}
\usepackage{booktabs,tabularx,array}
\newcolumntype{Y}{>{\raggedright\arraybackslash}X}
\usepackage[hidelinks]{hyperref}
\usetikzlibrary{positioning,arrows.meta}

\hypersetup{
  pdftitle={LiteSC: Lightweight Semantic Communication for Robust Wireless Telesurgical Video Transmission},
  pdfauthor={Zexin Deng, Zhenhui Yuan, Xiang Ji, Yi Han, and Gaofeng Li}
}

\newcommand{\method}{LiteSC}
\newcommand{\vect}[1]{\mathbf{#1}}

\begin{document}

\title{\fontsize{20}{24}\selectfont
LiteSC: Lightweight Semantic Communication for Robust\\
Wireless Telesurgical Video Transmission}

\author{%
Zexin Deng$^{1}$, Zhenhui Yuan$^{1}$, Xiang Ji$^{2}$, Yi Han$^{3}$, and Gaofeng Li$^{4}$%
\thanks{This work was supported by the Natural Science Foundation of Jiangxi Province under Grant 20262BAC240470 and by the EU Horizon Europe project DIRECT under Grant Agreement No. 101299316.}%
\thanks{$^{1}$Z. Deng and Z. Yuan are with the School of Engineering,
University of Warwick, Coventry CV4 7AL, United Kingdom
(e-mail: {zexin.deng, zhenhui.yuan}@warwick.ac.uk).
Z. Yuan also serves as a consultant member at the Institute for Applied
and Translational Technologies in Surgery, University Hospitals Coventry and
Warwickshire NHS Trust, Coventry, United Kingdom.}%
\thanks{$^{2}$X. Ji is with the School of Information Engineering,
Nanchang University, Nanchang 330031, China
(e-mail: tojixiang@ncu.edu.cn).}%
\thanks{$^{3}$Y. Han is with the School of Information Engineering,
Wuhan University of Technology, Wuhan, China
(e-mail: hanyi@whut.edu.cn).}%
\thanks{$^{4}$G. Li is with the College of Control Science and Engineering,
Zhejiang University, Hangzhou, China
(e-mail: gaofeng.li@zju.edu.cn).}%
}

\maketitle

\begin{abstract}
Reliable laparoscopic video is essential for telesurgery, yet fixed-rate
digital transmission can degrade abruptly under poor channel conditions.
We propose \method{}, a lightweight semantic communication framework
combining a frozen pretrained extractor, a compact joint source--channel
coding pair, and a receiver-adapted surgical renderer. The extractor produces
a latent with approximately 97.9\% fewer scalar values than the
red--green--blue (RGB) input, while inference requires neither segmentation
masks nor temporal reference frames. Using separate CholecSeg8k source videos
for training and testing, we evaluate \method{} over an additive white
Gaussian noise (AWGN) channel at symbol signal-to-noise ratios $E_s/N_0$ from
0 to 25~dB. At a channel bandwidth ratio of $\rho=0.0208$, equivalent to
0.0208 complex channel symbols per RGB source value, peak signal-to-noise
ratio (PSNR) rises from 28.7 to 31.9~dB, structural similarity index measure
(SSIM) from 0.869 to 0.919, and learned perceptual image patch similarity
(LPIPS) falls from 0.102 to 0.048. Against H.265 transmission protected by
low-density parity-check (LDPC) coding under the same channel-symbol budget,
\method{} achieves higher SSIM and lower LPIPS, although H.265+LDPC attains
a higher peak PSNR. These results show graceful laparoscopic frame
reconstruction under the evaluated AWGN conditions.
\end{abstract}


\section{Introduction}
\label{sec:intro}

\begin{table*}[!t]
\centering
\caption{Comparison with representative visual transmission frameworks.}
\label{tab:related_work_comparison}
\renewcommand{\arraystretch}{1.4}
\setlength{\tabcolsep}{4pt}

\resizebox{\textwidth}{!}{%
\begin{tabular}{|l|c|c|c|c|c|}
\hline
\textbf{Authors} &
\textbf{Pretrained Prior} &
\textbf{Receiver Adaptation} &
\textbf{Graceful Degradation} &
\textbf{Cue-Free Inference} &
\textbf{Scene (Dataset)} \\
\hline

Bourtsoulatze et al. (DeepJSCC)~
\cite{bourtsoulatze2019deepjscc}
& No
& No
& Yes
& Yes
& Natural images (CIFAR-10) \\
\hline

Tung and G\"und\"uz (DeepWiVe)~
\cite{tung2022deepwive}
& No
& No
& Yes
& No
& Human actions (UCF101) \\
\hline

Wang et al. (DVST)~
\cite{wang2023dvst}
& No
& No
& Yes
& No
& General video (HEVC, UVG) \\
\hline

Peng et al. (SCCVS)~
\cite{peng2026sccvs}
& No
& No
& Yes
& No
& Video surveillance (VIRAT) \\
\hline

Jiang et al. (Wireless SVC)~
\cite{jiang2023svc}
& No
& No
& No
& No
& Video conferencing (VoxCeleb) \\
\hline

Yang et al. (DiffJSCC)~
\cite{yang2025diffjscc}
& Yes
& Yes
& Yes
& No
& Image transmission (Kodak, ADE20K, CelebAHQ) \\
\hline

Guo et al. (DiSC-Med)~
\cite{guo2025discmed}
& No
& No
& Yes
& No
& Medical imaging (AMOS) \\
\hline

\textbf{\method{} (ours)}$^\ast$
& \textbf{Yes}
& \textbf{Yes}
& \textbf{Yes}
& \textbf{Yes}
& \textbf{Telesurgery (CholecSeg8k)} \\
\hline
\end{tabular}%
}

\vspace{4pt}
\parbox{\textwidth}{\scriptsize
$^\ast$\method{} combines a \textbf{pretrained perceptual prior},
\textbf{adaptation at the receiver}, and
\textbf{graceful degradation}, while requiring no auxiliary masks,
keypoints, captions, edge maps, or temporal reference frames during
inference. Pretrained prior refers to a pretrained representation or
reconstruction model rather than an auxiliary detector. A ``No'' under
graceful degradation indicates that this behavior was not established
in the cited evaluation. Scene reports the intended application and the
evaluation dataset; it does not imply live clinical deployment.
}
\end{table*}

Robot-assisted minimally invasive surgery (RAMIS) has achieved broad clinical
adoption, while telesurgery extends surgical operation across geographic
distance \cite{haidegger2022ramis,marescaux2001telesurgery}. In RAMIS-based
telesurgery, endoscopic video is a primary feedback channel for remote
teleoperation, making reliable visual delivery essential
\cite{anvari2005latency, tang2026lowlatency}. However, an uncompressed 8-bit red--green--blue
(RGB) stream at $1920\times1080$ resolution and 30~frames/s requires
approximately 1.49~Gbit/s. Although modern codecs such as H.265/High Efficiency
Video Coding (HEVC) substantially reduce this bitrate
\cite{sullivan2012hevc}, conventional digital transmission depends on
successful source and channel decoding and can exhibit abrupt quality
degradation near decoding thresholds, which is undesirable for continuous
surgical visual feedback.

Semantic communication and deep joint source--channel coding (DeepJSCC)
provide an alternative to conventional digital transmission by reducing
reliance on exact bit recovery and enabling graceful degradation as channel
conditions vary \cite{bourtsoulatze2019deepjscc}. Subsequent studies have
extended this principle to adaptive image transmission, wireless video, and
application-specific visual transmission
\cite{tung2022deepwive,wang2023dvst,
peng2026sccvs,jiang2023svc,yang2025diffjscc,guo2025discmed}.
Related telerobotic studies have investigated network-aware testbeds,
adaptive streaming, surgical video streaming, and multisensory feedback
\cite{deng2025telesim,deng2026tams,deng2026vista,
huang2025multisensory}.
Table~\ref{tab:related_work_comparison} summarizes representative visual
transmission frameworks and their key design characteristics.

Despite these advances, semantic and joint source--channel transmission of
laparoscopic video under varying wireless conditions remains insufficiently
explored.

Motivated by these gaps, our main contributions are summarized as follows.

\begin{itemize}
\setlength{\itemsep}{1pt}

\item We propose \method{}, a lightweight framewise semantic communication
framework for robust wireless laparoscopic video transmission, enabling
full-frame reconstruction without segmentation masks, keypoints, captions,
or temporal references during inference.

\item We establish a fixed pretrained latent interface at the transmitter,
decoupling visual representation extraction from wireless channel adaptation.
A compact joint source--channel coding (JSCC) pair protects this latent
representation, while the receiver-side renderer is adapted to laparoscopic
appearance. The latent contains approximately 97.9\% fewer scalar values than
the RGB input.

\item We evaluate \method{} over an additive white Gaussian noise (AWGN)
channel with $E_s/N_0$ from 0 to 25~dB under a matched communication budget
against H.265 baselines protected by low-density parity-check (LDPC) coding
using rates of $1/3$, $1/2$, and $2/3$ and 16-ary quadrature amplitude
modulation (16-QAM). \method{} exhibits gradual reconstruction-quality
variation with $E_s/N_0$ and achieves higher structural similarity and
substantially lower perceptual distance than the evaluated digital baselines.

\end{itemize}
\section{Proposed Method}
\label{sec:method}

\begin{figure*}[t]
\centering
\includegraphics[
    width=0.99\textwidth,
    trim=0 630pt 0 370pt,
    clip
]{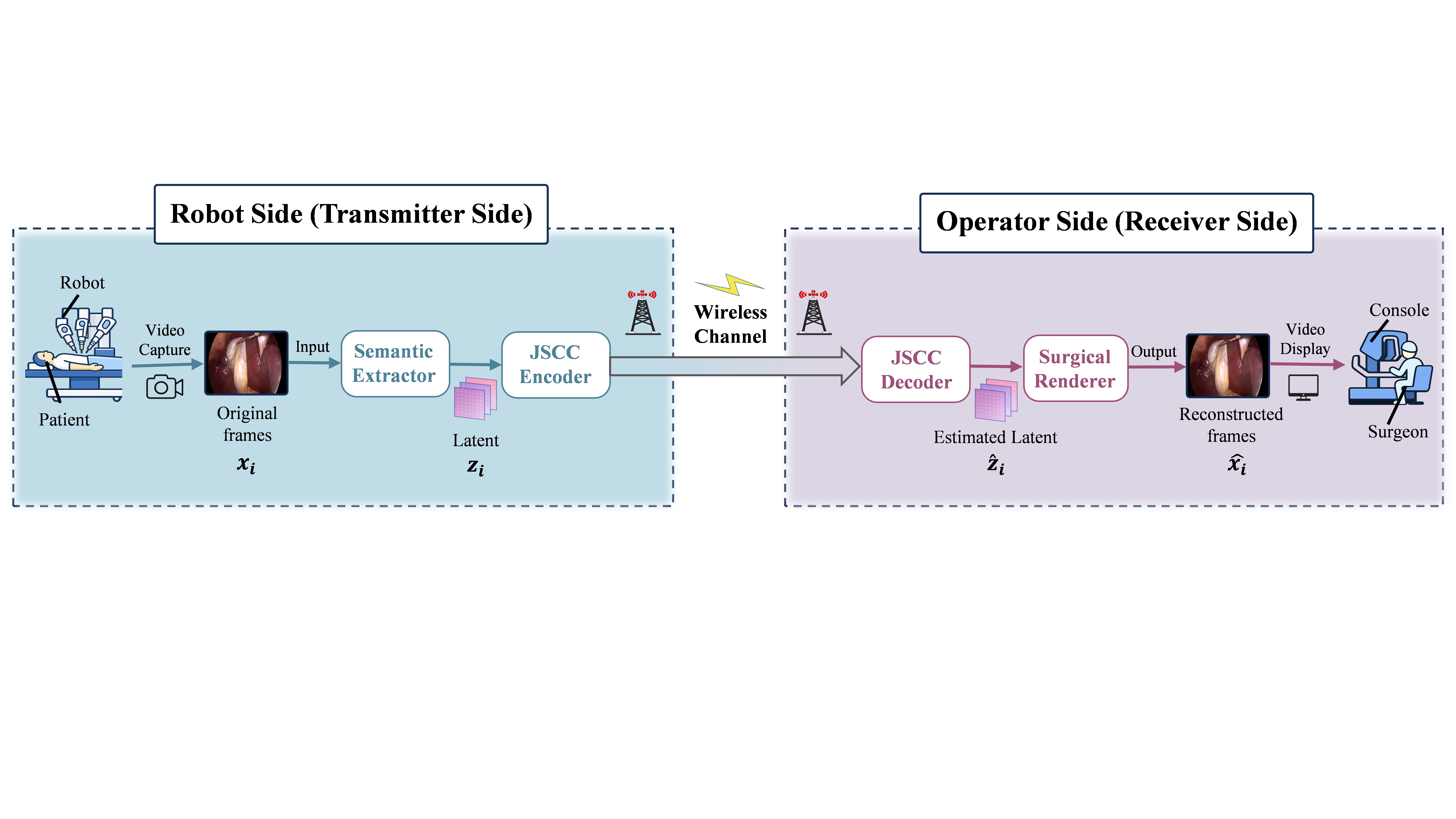}
\vspace{-3mm}
\caption{Overview of the proposed \method{} framework for wireless
telesurgical video transmission. The robot-side transmitter maps each
laparoscopic frame to a fixed pretrained latent representation and protects
it using the JSCC encoder. At the operator-side receiver, the JSCC decoder
estimates the latent representation, which is reconstructed by the adapted
surgical renderer for display to the surgeon.}
\label{fig:framework}
\end{figure*}

Figure~\ref{fig:framework} illustrates the architecture of \method{}.
This section first presents the transmission pipeline and then describes the
joint training of the JSCC pair and the receiver renderer.

\subsection{Transmission Pipeline}
\label{sec:pipeline}

The pipeline separates semantic extraction, channel protection, and surgical
reconstruction. Let
$\vect{x}_i\in[0,1]^{3\times H\times W}$ denote frame $i$. Each frame is
transmitted independently through the following four stages.

\subsubsection{Semantic Extraction}

We instantiate the semantic extractor with the Tiny AutoEncoder for Stable
Diffusion (TAESD)\footnote{TAESD:
\url{https://github.com/madebyollin/taesd/tree/main}}, a lightweight
autoencoder with 2.45 million parameters in total, distilled from the
autoencoder of latent diffusion models \cite{rombach2022ldm}. Its compact architecture
retains a compatible pretrained latent interface, and its encoder maps the
input frame to a perceptual latent with four channels:
\begin{equation}
\vect{z}_i
=
E_\varphi(\vect{x}_i)
\in
\mathbb{R}^{4\times h\times w},
\label{eq:enc}
\end{equation}
where $h$ and $w$ are the spatial dimensions produced by the extractor.
Relative to the RGB input, the reduction in scalar representation size is
\begin{equation}
r_{\mathrm{dim}}
=
\left(
1-\frac{4hw}{3HW}
\right)
\times 100\%
\approx
97.9\%.
\label{eq:latent_reduction}
\end{equation}

We retain the pretrained parameters of $E_\varphi$ throughout training. This
preserves a stable latent interface for channel protection and avoids
re-estimating the visual front end solely from the 6{,}080 surgical training
frames, while domain adaptation is placed at the receiver renderer.

\subsubsection{JSCC Encoding and Power Normalization}

The transmitter additionally contains a compact trainable JSCC encoder
$f_\theta$, which maps $\vect{z}_i$ to a real tensor
$\vect{u}_i\in\mathbb{R}^{C\times h\times w}$. A fixed pairing operator
$\mathcal{P}$ groups adjacent real values into a complex vector
$\vect{v}_i\in\mathbb{C}^{N}$, where $N=Chw/2$ and $C$ is even.

Following the average transmit power constraint commonly used in DeepJSCC
\cite{bourtsoulatze2019deepjscc}, we normalize the channel input as
$\vect{s}_i=\sqrt{N}\vect{v}_i/\|\vect{v}_i\|_2$, which satisfies
$\|\vect{s}_i\|_2^2/N=1$. The role of $f_\theta$ is therefore to protect the
fixed perceptual latent under the specified communication budget, rather than
to relearn the complete visual representation directly from pixels.

\subsubsection{Wireless Channel}

The normalized symbols traverse an AWGN channel:
\begin{equation}
\vect{y}_i
=
\vect{s}_i+\vect{n}_i,
\qquad
\vect{n}_i
\sim
\mathcal{CN}\!\left(
\vect{0},
10^{-\gamma_{\mathrm{dB}}/10}\vect{I}
\right),
\label{eq:chan}
\end{equation}
where $\gamma_{\mathrm{dB}}$ denotes the symbol signal-to-noise ratio (SNR),
$E_s/N_0$, in decibels, and $\mathcal{CN}$ denotes a circularly symmetric
complex Gaussian distribution. The channel layer introduces a continuous
perturbation and imposes no hard decoding decision.
Consequently, a decrease in $E_s/N_0$ does not by construction invalidate an
entire frame. Whether this architectural property produces graceful quality
degradation is evaluated empirically in Section~\ref{sec:results}.

\subsubsection{Latent Decoding and Surgical Rendering}

At the receiver, the JSCC decoder estimates the transmitted latent as
$\hat{\vect{z}}_i=g_\psi(\vect{y}_i)$, and the surgical renderer reconstructs
the frame as
$\hat{\vect{x}}_i=D_\omega(\hat{\vect{z}}_i)$. In the implementation, the
real and imaginary components of $\vect{y}_i$ are presented to $g_\psi$ as
separate real channels.

TAESD trades some fine reconstruction detail for a substantially smaller and
faster decoder, making the decoder a natural point for domain
adaptation. We therefore initialize $D_\omega$ from the pretrained decoder
$D_{\omega_0}$ and fine-tune it on surgical frames while keeping the extractor
unchanged. This preserves the compact latent interface at the transmitter while
specializing reconstruction of laparoscopic appearance, including tissue
texture, instrument structure, and specular highlights.

The reconstruction path requires no segmentation masks, keypoints, captions,
or previously reconstructed frames. This framewise design avoids dependence
on temporal state during inference, although it does not exploit redundancy
across frames.

The channel bandwidth ratio (CBR) is defined as the number of complex channel
symbols per RGB source value:
\begin{equation}
\rho
=
\frac{N}{3HW}
=
\frac{Chw}{6HW}
\approx
\frac{C}{384}.
\label{eq:cbr}
\end{equation}
For the evaluated dimensions, $C\in\{2,4,8\}$ gives
$\rho\in\{0.0052,0.0104,0.0208\}$ after rounding. Thus, $C$ controls the
communication budget without changing the dimensionality of the pretrained
perceptual latent.

\subsection{Training}
\label{sec:training}

We retain $\varphi$ unchanged, initialize $\omega$ from the pretrained decoder
parameters $\omega_0$, and jointly optimize $(\theta,\psi,\omega)$. At each
training step, $E_s/N_0$ is sampled through
$\gamma_{\mathrm{dB}}\sim\mathcal{U}(0,20)$, exposing each model to the
complete training interval.

The training objective combines the mean absolute pixel error with the learned
perceptual image patch similarity (LPIPS) distance:
\begin{equation}
\mathcal{L}
=
\mathbb{E}_{\vect{x},\gamma_{\mathrm{dB}},\vect{n}}
\left[
\frac{1}{3HW}
\|\hat{\vect{x}}-\vect{x}\|_1
+
\lambda_p
d_{\mathrm{LPIPS}}(\hat{\vect{x}},\vect{x})
\right],
\label{eq:loss}
\end{equation}
where $\vect{x}\sim\mathcal{D}_{\mathrm{train}}$,
$\gamma_{\mathrm{dB}}\sim\mathcal{U}(0,20)$, and the noise follows the model
in~\eqref{eq:chan}. The LPIPS distance follows \cite{zhang2018lpips}, and
$\lambda_p=1$.
The pixel term constrains pointwise reconstruction fidelity, while LPIPS
penalizes perceptually important differences that may not be adequately
represented by pixel error alone. For each value of $C$, the same trained
parameters are evaluated at all $E_s/N_0$ values; the model is not retrained
for individual channel conditions.

\section{Performance Evaluation}
\label{sec:experiments}

\begin{figure*}[!t]
\centering
\includegraphics[width=0.29\textwidth]{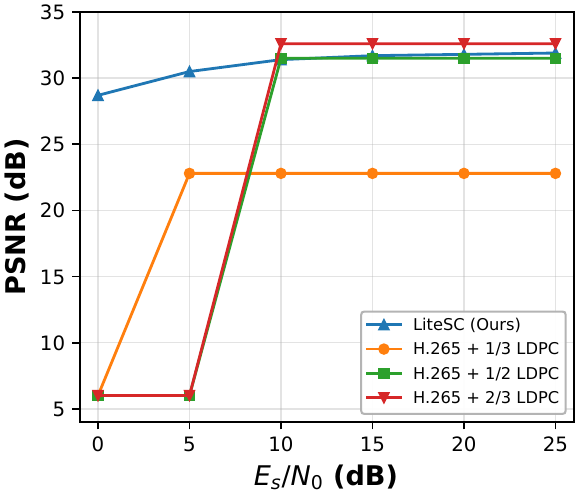}\hfill
\includegraphics[width=0.29\textwidth]{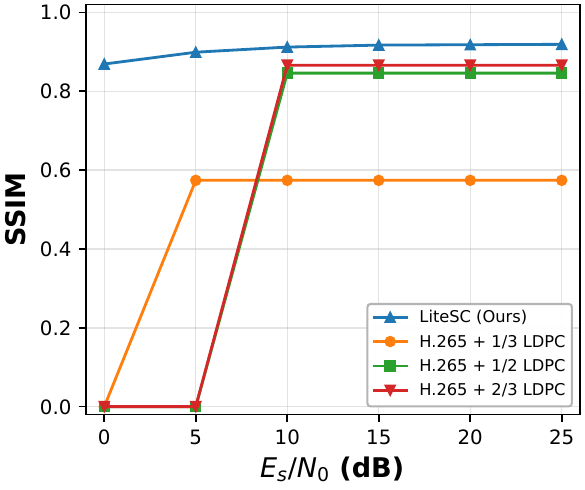}\hfill
\includegraphics[width=0.29\textwidth]{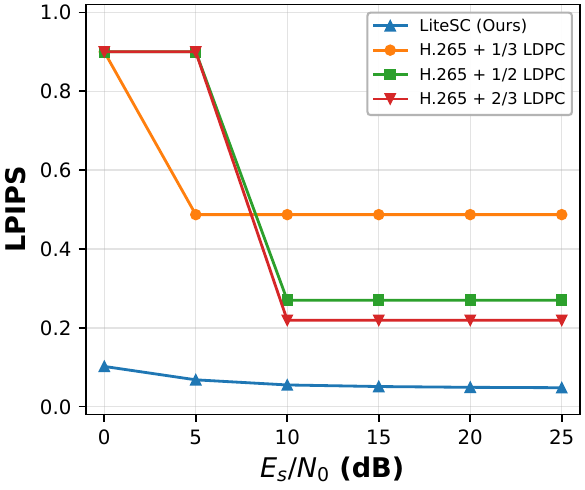}
\vspace{-3mm}
\caption{Performance over AWGN at $\rho=0.0208$. From left to right:
peak signal-to-noise ratio (PSNR), structural similarity index measure (SSIM),
and LPIPS versus $E_s/N_0$. \method{} exhibits gradual quality variation with
$E_s/N_0$, whereas the H.265+LDPC baselines show a sharper quality transition
as channel conditions improve.}
\label{fig:mainresults}
\end{figure*}

\subsection{Experimental Setup}
\label{sec:setup}
\subsubsection{Dataset and Preprocessing}
We evaluate \method{} on CholecSeg8k\footnote{CholecSeg8k:
\url{https://www.kaggle.com/datasets/newslab/cholecseg8k}}, which contains
8{,}080 annotated laparoscopic frames extracted from 17 video clips in
Cholec80 \cite{hong2020cholecseg8k}. To avoid overlap between frames from
the same source video, we split the dataset by video: 13 videos
(6{,}080 frames) are used for training and four videos (2{,}000 frames)
for testing. No source video contributes frames to both sets. All frames
are processed at their original resolution of $854\times480$.
\subsubsection{Training and Channel Settings}
The perceptual autoencoder is instantiated with TAESD. Its
encoder remains frozen, while the convolutional JSCC pair and the surgical
renderer are trained with AdamW. During training, $E_s/N_0$ is sampled from
$\gamma_{\mathrm{dB}}\sim\mathcal{U}(0,20)$.
Evaluation is performed over an AWGN channel at $E_s/N_0$ values in
$\{0,5,\ldots,25\}$~dB. For each frame and $E_s/N_0$ value, one noise realization is
generated using a fixed random seed. The main comparison uses a CBR of $\rho=0.0208$. Inference time is
measured with batch size one using 32-bit floating-point precision on an
NVIDIA RTX~4060 laptop graphics processing unit.
\subsubsection{Baseline Schemes}
For conventional video transmission, we compare \method{} with
H.265+LDPC baselines. H.265 is used for source compression, while LDPC
codes with coding rates of $1/3$, $1/2$, and $2/3$ are considered for
channel protection. All baseline transmissions employ 16-QAM modulation
and are evaluated at the same $E_s/N_0$ values under the same communication
budget.
\subsubsection{Evaluation Metrics}
Reconstruction quality is evaluated using PSNR, SSIM, and LPIPS.
PSNR measures pixel-level reconstruction fidelity, while SSIM evaluates
structural similarity between reconstructed and reference frames. LPIPS
measures perceptual similarity in a learned feature space using an AlexNet
backbone \cite{zhang2018lpips}. Higher PSNR and SSIM indicate better
reconstruction quality, whereas lower LPIPS is better.

\subsection{Quantitative Results}
\label{sec:results}
\subsubsection{Receiver Adaptation and Runtime}
With the pretrained autoencoder kept unchanged, the test frames achieve
31.9~dB PSNR, 0.917~SSIM, and 0.069~LPIPS, providing a reference for the
pretrained latent representation. In \method{}, the semantic extractor
remains fixed while the receiver renderer is adapted to laparoscopic images.
The extractor and renderer require 27 and 29~ms per frame, respectively,
giving a total model inference time of 56~ms. 
\subsubsection{Robustness Across Channel Conditions}
Figure~\ref{fig:mainresults} compares \method{} with the H.265+LDPC
baselines at $\rho=0.0208$. As $E_s/N_0$ increases from 0 to 25~dB,
\method{} improves gradually from 28.7 to 31.9~dB PSNR and from
0.869 to 0.919 SSIM, while LPIPS decreases from 0.102 to 0.048.
In contrast, the conventional digital baselines exhibit rate-dependent
decoding transitions followed by relatively stable plateaus at higher
$E_s/N_0$.
At 5~dB, \method{} achieves 30.5~dB PSNR, 0.899~SSIM, and 0.068~LPIPS,
whereas H.265+LDPC with rate $1/3$ reaches 22.8~dB PSNR, 0.574~SSIM,
and 0.487~LPIPS; the rate-$1/2$ and rate-$2/3$ baselines remain in the
low-quality decoding regime on average. By 10~dB, all three digital
baselines have crossed their observed decoding transition. The strongest
H.265+LDPC baseline, with rate $2/3$, reaches 32.6~dB PSNR, 0.866~SSIM,
and 0.219~LPIPS, while \method{} achieves approximately 31.4~dB PSNR,
0.912~SSIM, and 0.055~LPIPS. At 25~dB, \method{} reaches 31.9~dB PSNR,
0.919~SSIM, and 0.048~LPIPS, retaining higher structural and perceptual
fidelity despite the higher peak PSNR of the strongest digital baseline.
\begin{figure}[!t]
\centering
\includegraphics[
    width=\columnwidth,
    trim=10bp 100bp 10bp 35bp,
    clip
]{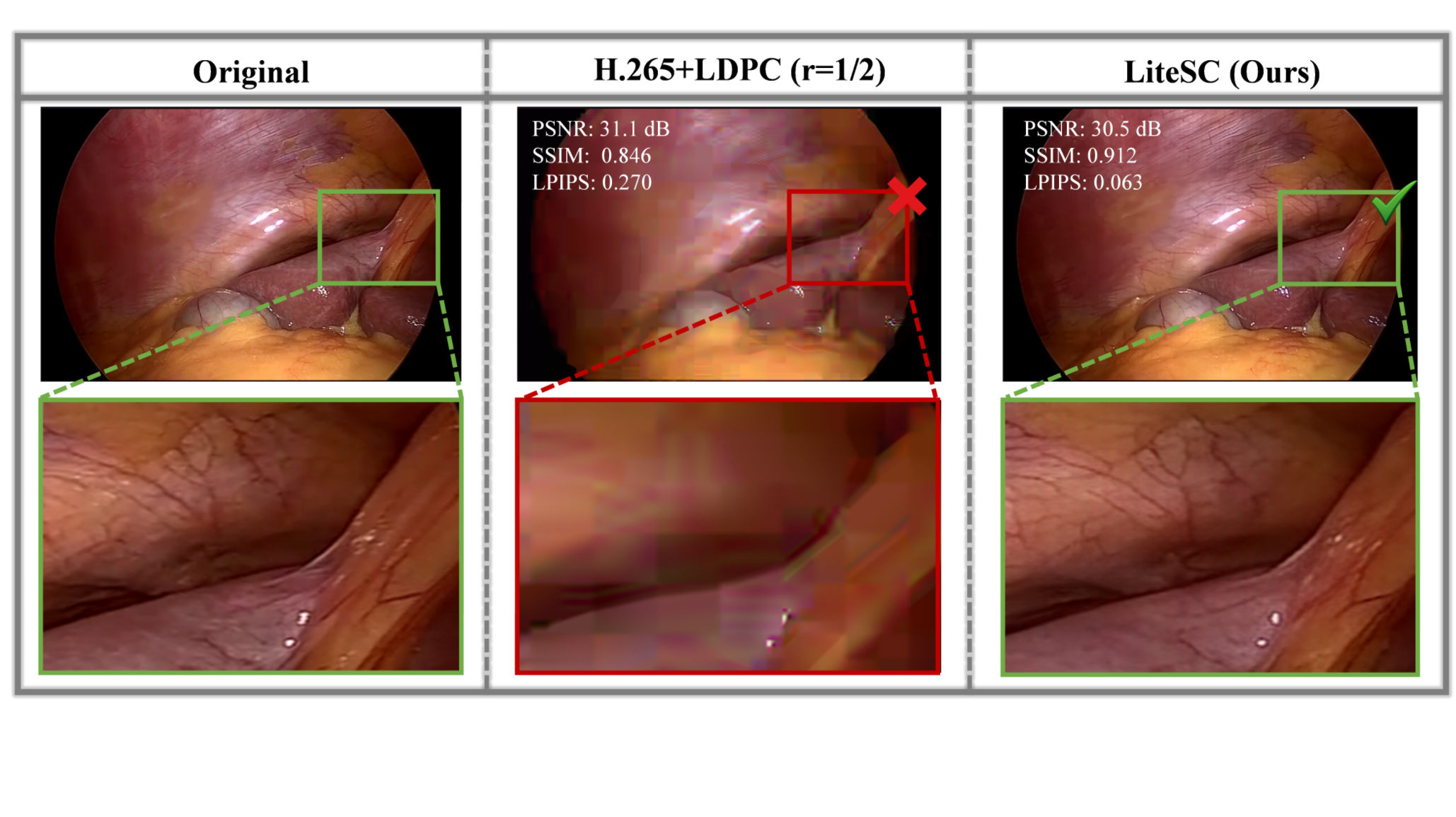}
\vspace{-6mm}
\caption{Qualitative comparison between \method{} and H.265+LDPC
($r=1/2$) on a held-out frame at 10~dB ($\rho=0.0208$).
Labels report per-frame PSNR, SSIM, and LPIPS.}
\label{fig:qual}
\end{figure}

\subsection{Qualitative Results}
\label{sec:qualitative}
Figure~\ref{fig:qual} presents a qualitative comparison on a held-out
laparoscopic frame at 10~dB and $\rho=0.0208$. In this example, the
H.265+LDPC ($r=1/2$) reconstruction attains slightly higher PSNR
(31.1 versus 30.5~dB), whereas
\method{} achieves higher SSIM (0.912 versus 0.846) and substantially lower
LPIPS (0.063 versus 0.270). The magnified region shows visible block
distortion and loss of fine tissue structure in the H.265+LDPC reconstruction,
while \method{} better preserves local tissue texture and structure.
This example illustrates that PSNR alone does not fully reflect perceptual
reconstruction quality under the constrained communication budget.

\section{Conclusion}
\label{sec:conclusion}

We propose \method{}, a lightweight framewise semantic communication framework
for robust wireless laparoscopic video transmission. By combining a fixed
pretrained latent interface, compact JSCC-based channel protection, and
receiver-side adaptation, \method{} supports cue-free full-frame
reconstruction under varying channel conditions. Over AWGN from 0 to 25~dB,
\method{} exhibits gradual quality variation, whereas the H.265+LDPC baselines
show sharper decoding transitions. \method{} further achieves higher SSIM and
substantially lower LPIPS despite a lower peak PSNR. 

\balance

\end{document}